\documentclass[prd,preprint,tightenlines,floatfix,showpacs,preprintnumbers,nofootinbib,eqsecnum]{revtex4}

 \usepackage[dvips,final]{graphicx}
  \usepackage{amssymb}
   \usepackage{amsmath}
    \usepackage{amsfonts}
     \usepackage{epsfig}
      \usepackage{bm} % bold math
\begin{document}

\title{Chiral vortical effect: black-hole vs. flat-space derivation}

%\author{A.I. Vasiliev}
%\email{khaidukov@itep.ru}
%\affiliation{ITEP, B. Cheremushkinskaya 25, Moscow, 117218 Russia}
\author{G. Yu. Prokhorov}
\email{prokhorov@theor.jinr.ru}
\affiliation{Joint Institute for Nuclear Research, Joliot-Curie str. 6, Dubna 141980, Russia}
\affiliation{Institute of Theoretical and Experimental Physics, NRC Kurchatov Institute, 
B. Cheremushkinskaya 25, Moscow 117218, Russia}
\author{O. V. Teryaev}
\email{teryaev@jinr.ru}
\affiliation{Joint Institute for Nuclear Research, Joliot-Curie str. 6, Dubna 141980, Russia}
\affiliation{Institute of Theoretical and Experimental Physics, NRC Kurchatov Institute,
 B. Cheremushkinskaya 25, Moscow 117218, Russia}
%\author{A.V. Sadofyev}
%\email{sadofyev@itep.ru}
\author{V. I. Zakharov}
\email{vzakharov@itep.ru}
\affiliation{Institute of Theoretical and Experimental Physics, NRC Kurchatov Institute, 
B. Cheremushkinskaya 25, Moscow 117218, Russia}
\affiliation{School of Biomedicine, Far Eastern Federal University, Vladivostok 690950, Russia}
%%
%\affiliation{Max-Planck Institut f\"{u}r Physik, 80805 M\"{u}nchen, Germany;}
%\affiliation{Moscow Inst Phys \& Technol, Dolgoprudny, Moscow Region, 141700 Russia.}

\begin{abstract}
The chiral vortical effect (CVE) was derived first for massless fermions,
within the framework of thermal quantum field theory.
%by Vilenkin in the year 1988. 
Recently, a dual description of the CVE,
as related to the radiation 
from the horizon of a rotating black hole was suggested. 
Generalizing the latter approach to the case of photons
we encounter a crucial factor of two difference with the 
predictions based on the thermal field theory. To elucidate the reason for this 
discrepancy we turn to the limit of  large spin $S$ of the massless particles. 
Within the gravitational approach the CVE grows as $S^3$ while the  %Kubo-type
flat-space
relations result in linear in $S$ dependence. We discuss also briefly an alternative
formulation of the presumed duality between the statistical and gravitational
approaches. 

\end{abstract}

\maketitle
%\section{Part I. Universal degrees of freedom in infrared}
%\subsection{Symmetries of action describing liquids}

\section{Gravitational chiral anomaly and CVE for spin-1/2 ~\centerline{particles}}

The chiral vortical effect (CVE)  is the flow of 
chirality of massless particles in a rotating medium along the vector 
of the angular velocity $\vec{\Omega}$. For a single right-handed Weyl fermion
one finds
\begin{equation}\label{one1}
\vec{j}^N_{CVE}~=~\left(\frac{\mu^2}{4\pi^2}+\frac{T^2}{12}\right) \vec{\Omega}~,
\end{equation} 
where $\vec{j}^N$ is the current of number of particles, $\mu$ is the chemical 
potential conjugated to the charge $Q^N$ associated with the current $j_{\mu}^N$,
$T$ is the temperature. In the pioneering paper \cite{vilenkin} Eq.~(\ref{one1}) 
is derived in the limit of non-interacting fermions by the statistical 
averaging of the matrix element of the current $j_{\mu}^N$.

A new era in theory of chiral effects began with the paper in Ref.
\cite{sonsurowka} which developed an approach 
valid in the strong-coupling limit,
or for ideal fluids. 
The basic idea is to rely only on the hydrodynamic expansion
and (anomalous) conservation laws.
Remarkably enough, 
Eq.~(\ref{one1}) survives the change in the framework.
Moreover, the coefficient in front of the $\mu^2\vec{\Omega}$
term
turns to be directly related to the coefficient
in the r.h.s. of the chiral anomaly
\begin{equation}\label{two1}
\partial^{\alpha}j_{\alpha}^N~=~-\frac{1}{32\pi^2}
\epsilon^{\alpha\beta\gamma\delta}F_{\alpha\beta}F_{\gamma\delta}~,
\end{equation}
where $F_{\alpha\beta}$ is the electromagnetic field strength tensor.
The coefficient in front of the $T^2$ term in Eq.~(\ref{one1}) 
remains undefined within the
hydrodynamic approach of Ref.~\cite{sonsurowka}  but can be fixed within the
framework of the 
thermal field theory \cite{songolkar,ren}.

The relation, if any, of the $T^2$-term in  Eq.~(\ref{one1}) 
to anomalies remained a kind 
of a mystery 
\footnote{For an earlier attempt to relate
the CVE to the gravitational anomaly see
\cite{megias}.}
until there appeared the paper \cite{stone}.
The main idea here goes back to the Refs.
\cite{wilzcek1, wilzcek2, wilzcek3, Banerjee:2007qs, Banerjee:2007uc, Banerjee:2008wq}
which relate the Hawking radiation, to the anomalies of the quantum field theory.
In more detail, it is suggested \cite{stone}
 to consider space-time with a boundary
imposed by the horizon of a rotating black hole.
Then one can check that near the horizon the r.h.s. of the gravitational chiral anomaly
\begin{equation}\label{four}
\nabla^{\alpha}j_{\alpha}^N~=~\frac{1}{768\pi^2\sqrt{-g}}
\epsilon^{\alpha\beta\gamma\delta}
R_{\alpha\beta\rho\sigma}R^{\rho\sigma}_{\gamma\delta}~,
\end{equation}
where $R_{\alpha\beta\gamma\delta}$ is the Riemann tensor,
is not vanishing.
Far off from the horizon,
where the r.h.s. of Eq.~(\ref{four}) vanishes,
there is a flow of chirality which can be found by integrating the r.h.s. of Eq.~(\ref{four})
\cite{stone}.

This asymptotic current coincides with
the $T^2\vec{\Omega}$ term in Eq.~(\ref{one})
{\it provided} that the generic temperature $T$ is replaced
by the Hawking temperature $T_H$ of the black hole
\begin{equation}\label{five}
T~\to~T_H~\equiv~\frac{a_H}{2\pi}~,
\end{equation}
and one keeps only the first term in the expansion in $\Omega$
(which is to be understood now as the angular velocity
at the horizon).
% We also note that there are a number of papers that consider
% chiral effects on the nontrivial geometrical background 
%\cite{Liu:2018xip,Flachi:2017vlp}.

Thus, in case of massless spin 1/2 particles there are two 
complementary ways of deriving the CVE, that is, within the statistical 
approach in  flat space
and in terms of black-hole physics. We are
considering generalization to the
photonic case and discuss the case of arbitrary spin. 
We obtain a quantitative prediction
 for the CVE of particles with arbitrary spin. Comparing this 
prediction with other approaches is a challenge for future research.

\section{Gravitational Chiral Anomaly and CVE for photons}

As is well known, the chirality of photons
is measured by the ``charge'' $\int d^3xK_0$
where the current $K^{\mu}$ is given by
\begin{equation}\label{current}
K^{\mu}~=~\frac{1}{\sqrt{-g}}\epsilon^{\mu\nu\rho\sigma}
A_{\nu}\partial_{\rho}A_{\sigma}~,
\end{equation}
where $A_{\mu}$ is the electromagnetic potential. The current (\ref{current})
is normalized in such a way that the corresponding (axial) charge
$Q^A_{photon}=\pm 1$
for the right- and left-hand polarized photons, respectively. 
Note also that the charge $Q^A_{photon}$
is gauge invariant, unlike the current itself.

The current (\ref{current}) is apparently not conserved,
since 
\begin{equation}\label{rhs}
\nabla_{\mu}K^{\mu}~\equiv~\frac{1}{4\sqrt{-g}} \epsilon^{\mu\nu\rho\sigma} F_{\mu\nu}F_{\rho\sigma}~.
\end{equation}
However, one can demonstrate \cite{vainshtein}
that naively  the expectation value 
of the r.h.s. of Eq.~(\ref{rhs}) for photons propagating in
external gravitational field vanishes, and
 there is analogy with the standard case
of charged massless fermions interacting with external electromagnetic field.
Moreover, there exists \cite{vainshtein,agullo,sadofyev1, Erdmenger:1999xx} 
the bosonic chiral gravitational anomaly
\begin{equation}\label{photonic}
<\nabla_{\mu}K^{\mu}>~=~\frac{1}{192\pi^2 \sqrt{-g}}\epsilon^{\alpha\beta\gamma\delta}
R_{\alpha\beta\rho\sigma}R^{\rho\sigma}_{\gamma\delta}~.
\end{equation}
Furthermore, Eq.~(\ref{photonic}) suffices to evaluate the chiral vortical effect 
for photons in terms of the black hole physics
following the logic of the paper  \cite{stone}.

Indeed, the chiral gravitational anomalies for spin-1/2 and spin-1 massless particles
are proportional to the same $R\tilde{R}$ and the effect of the
rotating black hole reduces to a universal geometric factor.
We are interested now in the spin dependence of the chiral vortical effect.
To elucidate  the spin dependence of the CVE
it is convenient to compare fermionic and bosonic fields with an equal number of 
chiral degrees of freedom,
that is normalize the photonic case to the
case of a Weyl spinor. By comparing Eqs.~(\ref{four}) and (\ref{photonic})
 we immediately
conclude
\begin{equation}\label{ratio}
\frac{(CVE)_{photons}}{(CVE)_{Weyl~fermions}}|_{black~ hole}~=~4~.
\end{equation} 
%In other words, a spin-squared dependence is suggested
%by this evaluation of the photonic vortical effect.
The problem is that the flat-space derivation suggests rather 
that the ratio (\ref{ratio})  is equal to 2, not 4.

\section{Evaluation of photonic CVE in flat space}
\subsection{ Kubo-type relation}
%In its original formulation, the chiral vortical effect is not related 
%to any anomaly, or 
%charge non-conservation. The vortical current (\ref{one1})  is conserved
%in flat space since
%$\mathrm{div}\,\vec{\Omega} = 0$.

The most common way to evaluate the CVE directly in flat space is to relate it
to a retarded, 3d Green's function using the technique 
\cite{megias} similar to
 Kubo relations. In more detail, define the conductivity 
$\sigma_{\Omega}$ as
\begin{equation}\label{sigma}
\vec{j}^N_{CVE}~=~\sigma_{\Omega}\vec{\Omega}~.
\end{equation}
Then, $\sigma_{\Omega}$ is
given by the retarded two-point Green’s function between the current
 $j^N_i$ and momentum density $T_{0j}$
 at zero frequency $\omega$ and small momenta $k_i$
\begin{equation}\label{retarded}
\lim_{k\to~0}{G_R(\omega,k)|_{\omega=0}}~=~i\epsilon_{ijn}k_n\sigma_{\Omega}~.
\end{equation}
Detailed calculations along these lines 
in case of charged spin-1/2 particles 
within thermal field theory can be found,
in particular, in  \cite{songolkar, ren}.

Eq.~(\ref{retarded}) can be generalized to
the case of photons \cite{songolkar}. 
%In fact the two-loop contribution to the CVE for spin-1/2 particles 
%factorizes into the product of one-loop chiral anomaly (\ref{two1})
%and of the CVE associated with the photonic current $K_{\mu}$.
The corresponding
conductivity, $\sigma_{\Omega}^{\gamma}$ for the current $K_{\mu}$
is expressed now in terms of the correlator between the
photonic current $K_i$  and the momentum density $T_{0j}$. 
The result of calculations can be 
 summarized as
\begin{equation}\label{ratio1}
\frac{(CVE)_{photons}}{(CVE)_{Weyl~spinor}}|_{Kubo~relation}~=~2~.
\end{equation}
Note that (\ref{ratio1}) differs from (\ref{ratio}) by a factor of 2.

Gauge invariance of
the results obtained remains a subtle point since the current  $K_{\mu}$
is not explicitly gauge invariant. 
Gauge invariance could be 
imposed explicitly by introducing
non-locality. In particular, for photons on mass-shell
the gauge-invariant current reads as:
\begin{equation}\label{kappa}
\kappa_{\mu}~=~(const)\frac{q_{\mu}}{q^2}F_{\alpha\beta}\tilde{F}^{\alpha\beta}~,
\end{equation}
where $q_{\mu}$ is the 4-momentum brought in by the current.
Note, however, that in the two cases most 
interesting for applications
the current $\kappa_{\mu}$ reduces in fact to $K_{\mu}$.
Namely, evaluation of the charge $\int d^3xK_0(x)$ assumes taking  the limit
$q_i= 0,~\omega\to 0$ in the language of the Fourier transform. In this limit
\begin{equation}
\lim_{q_i=0,\omega\to 0}{\kappa_0}~=~K_0~,
\end{equation}
and the  charge 
density $K_0$ understood as the component of the Fourier transform 
with $q_i\equiv 0, q_0\to 0$
turns gauge invariant.

Similarly, in case of the definition (\ref{retarded}), 
one considers the limit $\omega= 0, q_3\to 0$.
In this limit the non-local current $\kappa_{\mu}$ 
reduces to the component $K_3$
\begin{equation}
\lim_{\omega=0,q_3\to 0}{\kappa_3}~=~K_3~.
\end{equation}
In other words, the component of the Fourier transform of
the current defined by the Eq. (\ref{retarded}) is gauge invariant
and the conductivity $\sigma_{\Omega}$ is a physical observable.
%the same as used in the thermal field calculations. 
%Thus, the apparent lack of gauge invariance
%inherent to  using the polynomial form of the  current $K_{\mu}$
%seems actually not to be  a problem.
%might be not such a problem since in the limits considered the current
%$K_{\mu}$ coincides with non-local, explicitly gauge invariant currents. 

\subsection{CVE from the Sommerfeld expansion}

%The use of the Kubo relation (\ref{retarded}) reduces
%evaluation of  
%the conductivity $\sigma_{\Omega}$ to a one-loop perturbative
%calculation. As a result, the temperature dependence is factorized
%and the polynomial form of the current (\ref{sigma}) arises
%immediately. 

There is another approach to evaluate the CVE
by statistical averaging the matrix element of the corresponding current.
One finds first energy levels
as function of the momentum of the particles and 
of the angular velocity of the medium
and evaluates then the matrix element of the current for each mode.

The latter  technique was tried first
\cite{vilenkin}, with the following result for a Weyl fermion
\begin{equation}\label{nonperturbative}
{J}^N_{CVE}~=~\frac{1}{4\pi^2}
\int_{-\infty}^{\infty}\epsilon^2d\epsilon\cdot
\Big(\frac{1}{1+e^{\beta(\epsilon-(\mu+\Omega/2)}}-
\frac{1}{1+e^{\beta(\epsilon-(\mu-\Omega/2)}}\Big)
=\frac{\mu^2{\Omega}}{4\pi^2}+\frac{{\Omega}^3}{48\pi^2}+
\frac{T^2{\Omega}}{12}~.
\end{equation}
Note that 
%at the first stage, 
%when one derives the integral representation (\ref{nonperturbative})
%for the current, it is not obvious at all that the final result
%would be a polynomial. And 
there is no direct proof that 
the Kubo relation and the statistical averaging of the matrix element of
the current result in the same final expression. 
But the explicit calculations demonstrate consistency of the two 
approaches to evaluate the CVE for spin-1/2 particles.

As is emphasized recently \cite{stone}, 
the polynomiality  in
thermodynamic parameters of the Sommerfeld integrals (\ref{nonperturbative})
can be viewed
as an analogy to the field theoretic anomalies which 
are polynomial in the fields. Moreover, in case
of the $\mu^2\Omega$ term in the CVE current 
(\ref{nonperturbative}) the analogy can be traced 
algebraically on the level of Feynman graphs.
% using the substitution
%(\ref{three}).
%Note also that
%Eq.~(\ref{nonperturbative}) allows to analyze separately 
%contributions of the energies $\epsilon~\sim~T$ and $\epsilon~\sim~\Omega$,
%and this turns important in case of the photonic CVE, as we see in the
%next subsection.

%In case of the photonic CVE it would be natural to follow  as close as possible 
%the derivation of the fermionic CVE. And, indeed, 
%some steps in this direction were 
%made
%in \cite{vainshtein}. Namely, in the spinor-indices notations for the
%photonic field the free-field action takes on the form which is a close reminiscent
%of the action of a complex scalar field
%\begin{equation}
%S_{eff}~=~\int d^4x \bar{A}\nabla^2A~,
%\end{equation}
%where $\bar{A}\equiv A^{*}$ and under the chiral rotations
%$A~\to~e^{i\phi}A,~\bar{A}~\to~e^{-i\phi}\bar{A}$.
%Moreover, on mass-shell the fields $A,\bar{A}$ describe the 
%right- and left-handed photons. However, the fields $A,\bar{A}$
%are not Lorentz scalars and the corresponding propagators are
%gauge dependent. Thus, there is no direct analogy to the fermionic case.

Recently, the statistically averaged matrix element of the photonic current
$\vec{K}$ was evaluated in Ref.~\cite{landsteiner}. A complete set 
of solutions of the Maxwell equation satisfying the proper boundary conditions
was found as well as the energy levels. The final result 
%for the photonic CVE
is given by
\begin{eqnarray}\nonumber
&&J_{CVE}^{\gamma}~=~\frac{1}{8\pi^2}
\int_{\Omega^+}^{\infty}d\omega\int_{-\omega}^{\omega}
dk\big(\omega+k^2/\omega\big)\\
&&\Big(\frac{1}{e^{(\omega-\Omega)/T}-1}-
\frac{1}{e^{(\omega+\Omega)/T}-1}\Big) \label{landsteiner} 
=~\frac{2}{9}T^2\Omega+\mathcal{O}(\Omega^2)~,
\end{eqnarray}
where $\Omega^+=\Omega$ naively
but more realistically $\Omega^+$ satisfies the condition $\Omega^+>\Omega$.
In more detail, to avoid having velocity of rotation larger than the speed of light,
one introduces a
finite radius $R$ of the rotating cylinder and 
$\Omega^+~\to~\Omega$ in the limit $R\to\infty$. In this case, despite the fact that the Bose distribution in (\ref{landsteiner}) turns out to be singular in the limit $\Omega^+~\to~\Omega$, the integral (\ref{landsteiner}) is finite in this limit.

Note 
%that basic features of Eq.~(\ref{landsteiner}) are readily understood.
%Namely, the Fermi distribution of Eq.~(\ref{nonperturbative})
%is replaced by the Bose distribution and the energy shift  $\Omega/2$
%is replaced by $\Omega$ because of the change of spin of the particles
%from $S=1/2$ to $S=1$. On the other hand, to fix 
%the crucial overall factor one needs an explicit
%evaluation of the matrix elements of the current. Note 
that 
the final answer (\ref{landsteiner}) differs from  (\ref{ratio1})
by a factor of 4/3. 
The reason for that is not well understood.
We add some comments in the next subsection,
for further discussion see also \cite{landsteiner,mitkin1}.

To summarize, various approaches to evaluate the photonic
chiral vortical effect result in conflicting numerical values. 
This could be related to the subtleties of
the infrared regularization. We will discuss some aspects
of this problem next.

\subsection{Sensitivity to the infrared}
%Now we turn to discussion of another problem, indicated by 
%the Eqs.~(\ref{nonperturbative}), (\ref{landsteiner}).
Taken at face value Eqs.~(\ref{nonperturbative}), (\ref{landsteiner}) imply 
existence of negative modes at
$\epsilon \lesssim \Omega$. 
However, Eqs.~(\ref{nonperturbative}), (\ref{landsteiner}) themselves are derived 
under the assumption
that 
one can expand in $\Omega$ and this assumption is apparently not true at
$p\lesssim \Omega$.                                                               
In other words, the contribution of the region of small energies $\epsilon$
is to be considered more carefully.

The lowest levels in the rotating system can be 
 found by using the well-known analogy
between the magnetic field $\vec{H}$ and ``field of rotation'' 
$\vec{\Omega}$.
For massless charged fermions of spin 1/2 the Landau 
levels are given by
\begin{equation}\label{nielsen}
E_n~=~\pm\sqrt{2H(n+1/2)+P_3^2+H\sigma_3}~,
\end{equation} 
where $P_3$ is the momentum along the magnetic field and
 $\sigma_3=\pm 1$ 
is the sign of the spin projection
onto the direction of the magnetic field.
The lowest level $n=0$, $P_3=0$, $\sigma_3=-1$
is the famous zero model responsible for the chiral magnetic effect.
In case of the rotation, however, there is no zero mode for spin-1/2
\begin{equation}
E_n~>~0~,\quad spin~1/2~,  ~gravity~,
\end{equation}
since in the gravitational case
the gyromagnetic ratio is two times smaller than in case of the
electromagnetic interaction.

For spin-1 particle in rotating medium we would expect
that the zero mode comes back
\begin{equation}
E_{min}~=~0~,\quad spin~1~,~gravity~.
\end{equation}
 Indeed, the spin is doubled
and compensates for the just mentioned   loss of the factor of two. These expectations can be confronted 
with explicit evaluation
of the CVE effect for photons in the rotating medium,
see Eq.~(\ref{landsteiner}).
Clearly, Eq.~(\ref{landsteiner}) exhibits 
the zero mode which we are discussing. 
Keeping the radius $R$ of the rotating cylinder finite regularizes the expression 
for the current (\ref{landsteiner}) and eliminates the zero mode.
%However, it is not ruled out that the zero mode could reemerge,
%depending on further dynamical input. 
 %.,/

For higher spins, $  S\ge 3/2$, the lowest energy 
level is to be negative
\begin{equation}
E_{min}~<~0~,\quad S~\ge~3/2~, ~gravity~,
\end{equation}
and the perturbative vacuum is apparently unstable.

A careful analysis reveals further sources of infrared sensitivity.
In particular, it turns out that the chiral vortical current is model independent
only as far as it is evaluated on the axis of the rotation, or at
the radial coordinate $\rho~=~0$.
On the other hand, as we discussed above, 
see Eq.~(\ref{retarded}), evaluation of the CVE assumes that
the momentum tends to zero, $q_i~\to~0$. And it is only in this limit
that we have gauge invariance granted. The conditions $\rho=0$ and $q_i\to~0$
are at least formally in conflict with each other. In principle, 
this could be a source of discrepancy
between results of various evaluations of the photonic CVE.

Another typical example of an infrared problem
is described in Ref.~\cite{adler}. One introduces a massless, 
spin-3/2 charged field $\psi_{\nu}$. To avoid the infinite radiation, 
one adds a  massless
spin-1/2 field, $\lambda$,  along with interaction which generates mass
\begin{equation}
S_{m}~
=~m\int d^4x
\big(\bar{\lambda}\gamma^{\nu}\psi_{\nu}-\bar{\psi}_{\nu}\gamma^{\nu}\lambda\big)~.
\end{equation} 
This solves the infrared problem but modifies also the chiral anomaly \cite{adler}
because of the introduction of an extra particle $\lambda$.

To summarize, there are
various sources of sensitivity of the global picture to details of the 
dynamics in the infrared. However, the high temperature contribution 
to the CVE 
might well be protected against these infrared sensitive effects. 
Indeed, there is no reason to expect that the two regions,
infrared-sensitive and high-temperature ones, give
parametrically similar results.

Under this assumption, we proceed to consider the higher-spin case.

\subsection{Limit of large spin of massless particles}

%In this subsection, we argue that in the large-spin limit 
% the disagreement between the two ways of 
%evaluating
%the chiral vortical effect becomes qualitative.

The coefficient in front of the
gravitational anomaly grows with spin $S$ of the massless chiral particles as $S^3$
for large $S$ \cite{duff1, duff2, vainshtein1}
\begin{eqnarray}
\nabla_{\mu} K^{\mu}_S=
\frac{(-1)^{2S}(2S^3-S)}{192 \pi^2\sqrt{-g}}\epsilon^{\mu\nu\rho\sigma}
R_{\mu\nu\kappa\lambda}{R_{\rho\sigma}}^{\kappa\lambda}~,
\label{anom spin S}
\end{eqnarray}
where $K_S^{\mu}$ is the chiral
current for massless particles of spin $S$, an analog of the current $K^{\mu}$ in the
photonic case. The current $K^{\mu}_S$ can explicitly be constructed in terms
of the Pauli-Lubanski pseudovector \cite{vainshtein1}. 
Basing  on the general formula (\ref{anom spin S}) and following 
\cite{stone}, we predict the $ T^2 $ term in the vortical current
\begin{equation}\label{predict S}
\vec{K}_{S}= \frac{(-1)^{2S}(2S^3-S)}{3} T^2\vec{\Omega}~,\quad (gravitational~anomaly)~.
\end{equation}

A striking feature of 
the prediction (\ref{predict S}) 
for the CVE of particles with large spin $S$
is its $S^3$ dependence. This spin dependence is not reproduced 
within thermal field theory in flat space.
Indeed,
 within the thermal field theory
 the only source of the growth of the vortical current 
with spin $S$
is the effective coupling
 $\vec{S}\cdot \vec{\Omega}$, specific for the 
equilibrium. Thus, the term linear in $\Omega$ grows also linearly
in spin $S$. These expectations are confirmed by explicit calculations,
see, in particular \cite{sadofyev2} and references therein.
According to \cite{sadofyev2} the flow of chirality $J_{\chi}$ carried by the spin-$S$
massless particles is given by
\begin{equation}\label{sadofyev}
\vec{J}_{\chi}~=~\frac{{S}\cdot\vec{\Omega}}{\pi^2}\sum_{\pm}
\int_0^{\infty}f_{\pm}(p)pdp~,
\end{equation}
where the summation is over the chiral states
with projection on the momentum equal to $\pm S$, $f_{\pm}(p)$
are the Bose or Fermi distributions, whichever relevant, and
one reserves for non-vanishing chemical potentials $\mu_{\pm}$.

%To our mind, however, the best motivated estimate of
%the vortical current $\vec{K}_S$ for an arbitrary  spin $S$
%reads:
%\begin{eqnarray}\label{temperature}
%\vec{K}_S~=~\frac{4S}{2S+1}\frac{T^2}{12}\vec{\Omega}\qquad half-integer~S~,
%\nonumber \\
%\vec{K}_S~=~\frac{8S}{2S+1}\frac{T^2}{12}\vec{\Omega}\qquad integer~S~.
%\end{eqnarray} 
%Here the extra factor $2/(2S+1)$ generalizes the factor 2/3 established
%first in Ref.~\cite{landsteiner} in the photonic case, see Eq.
%(\ref{landsteiner}).

To summarize, consideration of higher spins elevates the 
mismatch between the two ways of evaluating the CVE 
to a qualitative effect.

\subsection{Spin dependence of vortical current within the  gravitational approach}

The predictions (\ref{predict S}) and (\ref{sadofyev})
differ qualitatively. The origin of the disagreement is that
%in thermal field theory
in flat space
the effect is controlled primarily
by the number of degrees of freedom which for massless particles does not grow with
their spin $S$. The strong $S$ dependence in case of the gravitational anomaly
is apparently due to a specific gravitational interaction with spin.

%Note that if such a strong dependence on $S$ 
%were manifested in the effect linear in acceleration $a$,
%this would be in contradiction with the equivalence principle. However,
%we are discussing now the vortical current which is quadratic in the acceleration $a$.

As is noticed in Refs.~\cite{duff1, duff2}, the $S^3$ dependence 
of the gravitational anomaly goes back to the coupling of the Riemann tensor
to the spin of the particles
\begin{equation}\label{quadropole}
X~=~\Sigma^{\mu\nu}R_{\mu\nu}^{~~ab}\Sigma_{ab}~,
\end{equation}
where $\Sigma^{\mu\nu}$ is the spin operator. 
%with covariant, respectively local Lorentzian indices. 
 Physically, this coupling corresponds to the
interaction of the gravitational wave with the quadrupole moment of the particle
\footnote{We are grateful to A.I. Vainshtein for this remark.}. 

In the next section we describe briefly another formulation of the statistical
approach which provides us with another derivation of the $S^3$ dependence.

\section{Temperature-acceleration, $T\leftrightarrow \frac{a}{2\pi},$ Duality?}

\subsection{Derivation of the Unruh temperature within statistical approach} 

Discovery of the Unruh temperature, or temperature seen by an accelerated observer
\begin{equation}\label{unruh}
T_U~=~\frac{a}{2\pi}~,
\end{equation}
 established for the first time a kind of equivalence, or duality between
acceleration $a$ and the temperature $T$.
The next step was made in Refs.~\cite{wilzcek1,wilzcek2,wilzcek3,Banerjee:2007qs,Banerjee:2007uc,
Banerjee:2008wq}. Here, the basic idea is that at the horizon of the black hole
there is an intrinsic right-left asymmetry
since  the particles are emitted outwards of the horizon and absorbed inwards.
The precise measure of the particle production is provided by the 
quantum anomalies in terms of the gravitational field, or acceleration on the horizon, 
$a_H$.
By matching the flow of the particles far off from the horizon 
to the radiation from a black body
one is rederiving the Hawking temperature $T_H$
\begin{equation}
T_H~=~\frac{a_H}{2\pi}~.
\end{equation}
Finally, Ref.~\cite{stone} suggested a similar construction to relate the chiral
 vortical effect
in flat space to the  radiation from a rotating black brane.

In all these cases one and the same phenomenon is described either 
in terms of the statistical theory as a process in equilibrium, or in 
terms of quantum field theory, as
a manifestation of an anomaly in external gravitational field.

There exists another systematic way to 
relate physics in the equilibrium in flat space to the physics 
in external gravitational field 
and vice verse, for introduction and further references 
see \cite{becattini}. On the statistical side,
the crucial point is the introduction of the following density operator $\hat{\rho}$
\begin{equation}\label{two3}
\hat{\rho}~=~\frac{1}{Z}\exp\Big(-\beta_{\mu}\hat{P}^{\mu}+
\frac{1}{2}\varpi_{\mu\nu}\hat{J}^{\mu\nu}+\xi \hat{Q}\Big)~,
\end{equation}
where $  \xi=\mu/T$,
$\hat{P}_{\mu}$ is the  4-momentum, 
$\hat{J}^{\mu\nu}$ are generators of the Lorentz
transformations and  $\hat{Q}$ is a conserved charge.
Moreover,
%\begin{equation}\label{two}
$\varpi_{\mu\nu}~=~{1}/{2}\big(\partial_{\nu}\beta_{\mu}-
\partial_{\mu}\beta_{\nu}\big)~$
%\end{equation}
is the tensor of the thermal vorticity, and
$\beta_{\mu} =u_{\mu}/T$.
The operator $\hat{J}^{\mu\nu}$ in Eq.~(\ref{two3}) 
can be rewritten as
\begin{equation}\label{one}
\hat{J}^{\mu\nu}~=~u^{\mu}\hat{K}^{\nu}-u^{\nu}\hat{K}^{\mu}-
\epsilon^{\mu\nu\rho\sigma}u_{\rho}\hat{J}_{\sigma}~,
\end{equation}
where $\hat{K}^{\mu}$ is the boost operator and  $\hat{J}_{\nu}$
is the operator of angular momentum.  
%The meaning of Eq.~(\ref{one}) is easier to appreciate
%upon introduction of the 4-vectors of (thermal) acceleration,
%$\alpha_{\mu}$ and of (thermal) vorticity, $w_{\mu}$
Furthermore, it is useful to introduce 4-vectors $\alpha_{\mu}, w_{\mu}$
\begin{equation}
\alpha_{\mu}~=~\varpi_{\mu\nu}u^{\nu}~,\qquad w_{\mu}~=~(1/2)
\epsilon_{\mu\nu\alpha\beta}u^{\nu}\varpi^{\alpha\beta}~.
\end{equation}
In the rest frame
 $T\cdot \alpha^i$ and $T\cdot w^i$
coincide with the standard 3-vectors of acceleration $\vec{a}$ and 
of the angular velocity $\vec{\Omega}$,
respectively. 

A special case of the operator $ \hat{\rho} $ is better known, when it describes a medium with a finite constant angular velocity $ \vec{\Omega} $ 
%and projection of the angular momentum to the axis of rotation $ \hat{J}_z $ 
\cite{becattini, Ambrus:2019ayb}
\begin{equation}
\hat{\rho}~=~\frac{1}{Z}\exp\Big(-\hat{H}/T_0+
\Omega\hat{J}_z/T_0+\mu_0 \hat{Q}/T_0\Big)~,
\end{equation}
where the indices $ 0 $ indicate that the quantities are measured in a stationary reference frame.

%In an alternative form
%\begin{equation}\label{three1}
%\varpi_{\mu\nu}~=~\epsilon_{\mu\nu\alpha\beta}w^{\alpha} u^{\beta}+
%\alpha_{\mu}u_{\nu}-\alpha_{\nu}u_{\mu}~.
%\end{equation}
%Eqs.~(\ref{two3}),  (\ref{two}), (\ref{three1}) 
%allow to consider the case when both acceleration and 
%rotation, and temperature  are treated as independent parameters.

%The form of the density operator (\ref{two3}), 
%built on the generators of the Poincar\'e group
%is deeply rooted in symmetries 
%of the Minkowskian space-time.
%One can say, therefore, that
%the equilibrium in rotating and accelerated medium, 
%with constant $\varpi_{\mu\nu} $
%represents the most general case intrinsic to  the statistical approach,
%and it cannot be extended further. 

Note that 
the boost operators $\hat{K}^i$, although conserved, do not commute
with the Hamiltonian $\hat{H}$
\begin{equation}\label{conservation}
i\frac{d\hat{K}^i}{dt}~=~0~,
%i\frac{\partial \hat{K}^i}{\partial t}+[\hat{K}^i,\hat{H]}~=~0~, \nonumber \\ 
\qquad [\hat{K}^i,\hat{H}]~=~-i\hat{P}^i~,
\end{equation}
and the equations (\ref{conservation}) are consistent with each other
because of an explicit dependence of $\hat{K}^i$ on time \cite{becattini}.
For this reason actual calculations based 
on the density operator (\ref{two3}) in case of $\vec{a}\neq 0$
are highly non-trivial and the corresponding technique
has been worked out
only recently \cite{becattini}.

Applications of this technique have turned successful.
In particular, the Unruh temperature can be defined now 
\cite{becattini} in terms of the energy density $T_{00}$ as a function 
of acceleration $a$ and of the temperature, $T_{00}(a,T)$
\begin{equation}\label{condition}
T_{00}(a,T)|_{T=T_{U}}~=~0~.
\end{equation}
It was demonstrated \cite{prokhorov5} that in case 
of massless spin-1/2 and spin-0 particles
 the condition (\ref{condition}) allows to determine the Unruh 
temperature 
%in a closedform, 
without introducing any parameter or subtractions. 
%Moreover,
%there is a condition 
%\begin{equation}
%T~\ge~T_U~,
%\end{equation}
%otherwise, the vacuum is unstable.

Note that within the approach (\ref{two3})
acceleration and temperature are treated as independent variables.
To match this freedom on the gravitational side
one considers \cite{prokhorov5} the Rindler space with a conical singularity 
and defines the acceleration
and temperature geometrically. 
The duality works perfectly well \cite{prokhorov5}. One can say that 
the case $\Omega=0,~(a,T)~\neq 0$ is fully understood in terms of the
duality between gravitational and statistical approaches. Here we are considering
both $a$ and $\Omega$ non-vanishing and it is a generic feature that the formalism
is becoming much more complicated, see in particular \cite{martinez}.

\subsection{Coupling  of acceleration and vorticity to the spin}

The density operator (\ref{two3}) which we are exploiting, 
 introduces
the following  effective interaction, specific for the physics of equilibrium
\begin{equation}
\delta \hat{H}~=~-\vec{\Omega}\cdot\vec{\hat{J}}-\vec{a}\cdot\vec{\hat{K}}~, \label{eff}
\end{equation}
%where $\vec{\hat{J}}$ and $\vec{\hat{K}}$ are  generators of rotations and boosts. In particular, 
In the nonrelativistic limit the part with boost leads to energy of a particle in a constant gravitational field $ a $ \cite{becattini}
\begin{equation}
 \hat{H}-a \hat{K}_z~=~mc^2 +\hat{p}^2/2m-m a \hat{z}~. \label{eff1}
\end{equation}
More generally, the terms with boost and angular momentum provide that the evolution operator generates a transition between instantaneous inertial rest frames at different time moments, since these transformations in the presence of acceleration are not reduced only to time translations, and also include Lorentz transformations \cite{Korsbakken:2004bv}.

We consider the case  of the vectors $\vec{\Omega}$ and $\vec{a}$
parallel to each other, $\vec{\Omega}~||~\vec{a}$. Then it is straightforward
to evaluate the correction to the energy of a chiral state of spin-1/2 polarized
along the vector $\vec{\Omega},~\vec{a}$
\begin{equation}
\delta E~=~-(\Omega_z-ia_z)\sigma_z/2~.
\end{equation}
The only non-trivial point here is the emergence of an ``imaginary energy" $\delta E \sim ia/2$.
The reason for that is the lack of a unitary representation for the boosts operator in case of 
a chiral multiplet.
As a result, one uses an anti-unitary realization of the boost operator,
$K^i~\sim~i\sigma^i$ which, however, respects commutation relations
between the operators.
Thermodynamically, this  imaginary energy does not make
any trouble since the odd powers of $a$ are canceled out between 
contributions of particles 
and anti-particles,
while even powers of $a$ survive. 

Now, the generalization to the case of massless particles of higher spin $S$ is trivial
\begin{equation}
\sigma_z/2~\to~S_z~, \quad \delta E~\to~(\vec{\Omega}-i\vec{a})\cdot \vec{S}~.
\end{equation} 
Estimates of spin dependences of other observables is also straightforward. Namely,
each power of $a$ or $\Omega$ is accompanied by a factor of $S$,
while the temperature is ``unaware" of the spin since there are two degrees of freedom 
for any massless particle with $S\neq 0$. 

Consider as 
an example the chiral vortical effect.
On the dimensional grounds
\begin{equation}\label{estimate}
J_{CVE}~\sim~c_1 T^2\Omega+ c_2 a^2\Omega~,
\end{equation}
where the coefficients $ c_1 $ and $ c_2 $ are to be calculated.
Following the logic outlined above, we come to the estimates
\begin{equation}
c_1~\sim~S~,\quad c_2~\sim~S^3~.
\end{equation}
 The $S^3$ dependence 
of $c_2$ matches perfectly
the prediction (\ref{predict S})
based on the gravitational anomaly.

To summarize, straightforward estimates within the statistical approach based on the
density operator (\ref{two3}) immediately result in a $S^3$ dependence of the
chiral vortical effect for higher-spin particles. 

\section{Conclusions}

Two approaches have been tried 
to evaluate the chiral vortical effect  (CVE) for massless spin-1/2 particles,
with identical results obtained.
The first approach is based on statistical averaging of the matrix elements of
the axial current. The other one relates the CVE to  the chiral gravitational anomaly.
The latter approach can immediately be generalized to the case
of arbitrary spin of particles interacting with external gravitational field.

The prediction for the photonic CVE obtained in this way can be compared
with the    results obtained within the statistical approach. In particular,
application of the Kubo-type relation to evaluate the photonic CVE gives
an answer \cite{songolkar,ren} which differs by a factor of two from 
the 
gravitational approach. However, there is no consensus yet in
the literature concerning the value of the vortical conductivity $\sigma^{\gamma}_{\Omega}$.
The source of uncertainty is apparently dependence on the regularization procedure in
the infrared.
Thus, it is not ruled out that the final result for photons would agree with the prediction
of the gravitational approach.

To have a clearer case we turn to consideration of the limit of 
large spin $S$ of the constituents. In this case the predictions for the vortical conductivity
differ qualitatively
\begin{eqnarray}
&&\sigma_{\Omega}~\sim~S^3\quad (gravitational ~anomaly)~,\\\nonumber
&&\sigma_{\Omega}~\sim~S \quad (thermal~ field~ theory)~.
\end{eqnarray}
We also try an alternative statistical approach
based on the use of the density operator (\ref{two3}). 
 This version of the statistical approach
does reproduce the $S^3$ dependence 
of the vortical effect in accelerated medium.
The basic element of the derivation is demonstration of existence of 
coupling of acceleration to spin, $\delta L~\sim~\vec{a}\cdot\vec{S}$,
specific for physics of the equilibrium \footnote{Similar coupling appears in the case of the Coriolis force, which is associated with a term in the Lagrangian of the form $ \delta L= m \vec{v} \cdot  \vec{\Omega} \times \vec{r} =  \vec{\Omega} \cdot \vec{J} $, while the interchange of spin and orbital momenta is
justified by equivalence principle.}.
On the other hand, linear in the acceleration
terms drop off thermodynamically, and there is no contradiction with the 
equivalence principle for this reason.  

%We emphasize, however, that in order to reproduce the cubic dependence $ S^3 $ from (\ref{estimate}) we would have to use the  duality (\ref{unruh}) arising at the quantum-field level. Considering the kinematic relationship, say, centripetal acceleration and angular velocity, according to which $ a=\Omega^2r $ is necessary, when considering the dependence of this effect on distance (see \cite{vilenkin}) and is not associated with a change in the coefficient in front of the term $ T^2 $.

There are many questions left open.
The main reservation concerning the status of the results obtained
is that theories involving massless particle with large spin $S$ have intrinsic 
problems and difficulties. 

\section{Acknowledgments}
The authors are thankful to P. G. Mitkin, A. V. Sadofyev, A. I. Vainshtein 
and A. I. Vasiliev for valuable discussions and to J. Erdmenger and R. Banerjee
for communications. The work was supported by 
the RFBR grant No. 18-02-40056.

\end{document}